\documentclass[conference, letterpaper]{IEEEtran}

\ifCLASSINFOpdf
\else
\fi

\ifCLASSINFOpdf
   \usepackage[pdftex]{graphicx}
\else
\fi

\usepackage{amsmath,amsfonts,amssymb}%[cmex10]
\usepackage{color}

\usepackage{fancyhdr}

\renewcommand{\thispagestyle}[2]{}

\fancypagestyle{plain}{
        \fancyhead{}
        \fancyhead[C]{first page center header}
        \fancyfoot{}
        \fancyfoot[C]{first page center footer}
}
\begin{document}

\newtheorem{theorem}{Theorem}
\newtheorem{proposition}[theorem]{Proposition}
\newtheorem{definition}[theorem]{Definition}
\newtheorem{remark}[theorem]{Remark}
\newtheorem{example}[theorem]{Example}
\newtheorem{claim}[theorem]{Claim}
\newtheorem{problem}[theorem]{Problem}
\newtheorem{corollary}[theorem]{Corollary}

\title{Solutions of partition function-based TU games\\ for cooperative communication networking}

\author{\IEEEauthorblockN{Giovanni Rossi}
\IEEEauthorblockA{Dept of Computer Science and Engineering - DISI\\
University of Bologna\\
40126 Bologna, Italy\\
Email: giovanni.rossi6@unibo.it}}

\maketitle

\begin{abstract}
In networked communications nodes choose among available actions and benefit from exchanging information through edges, while continuous technological progress fosters system functionings that increasingly often rely on cooperation. Growing attention is being placed on coalition formation, where each node chooses what coalition to join, while the surplus generated by cooperation is an amount of TU (transferable utility) quantified by a real-valued function defined on partitions -or even embedded coalitions- of nodes. A TU-sharing rule is thus essential, as how players are rewarded determines their behavior. This work offers a new option for distributing partition function-based surpluses, dealing with cooperative game theory in terms of both global games and games in partition function form, namely lattice functions, while the sharing rule is a point-valued solution or value. The novelty is grounded on the combinatorial definition of such solutions as lattice functions whose M\"obius inversion lives only on atoms, i.e. on the first level of the lattice. While simply rephrasing the traditional solution concept for standard coalitional games, this leads to distribute the surplus generated by partitions across the edges of the network, as the atoms among partitions are unordered pairs of players. These shares of edges are further divided between nodes, but the corresponding
Shapley value is very different from the traditional one and leads to two alternative forms, obtained by focusing either on
marginal contributions along maximal chains, or else on the uniform division of Harsanyi dividends. The core is also addressed, and supermodularity is no longer sufficient for its non-emptiness. 
\end{abstract}


\section{Introduction}
For the present purposes, a communication network may be looked at as a simple graph $G^t=(N,E^t)$ varying in time $t$. The set $N=\{1,\ldots,n\}$ of nodes or vertices can be
constant, while the set $E^t\subseteq N_2=\{\{i,j\}:1\leq i<j\leq n\}$ of edges in the network at any time $t$ results from node behavior for given technological environment.
In fact, in game-theoretical models of communication networking \cite{GameTheoryCommunicationb00k2012,CoalitionalGameThoeryTutorial} nodes are players whose behavior generally
realizes as a time-sequence of choices over available actions. Most importantly, these players attain a utility by exchanging information with each other through the network,
while the functioning of this latter incerasingly often relies on cooperation. In this view, \textit{``cooperative game theory provides a variety of tools useful in many
applications''}, allowing \textit{``to model a broad range of problems, including cooperative behavior, fairness in cooperation, network formation, cooperative strategies, and
incentives for cooperation''}, and with applications such as \textit{``information trust management in wireless networks, multi-hop cognitive radio, relay selection in cooperative
communication, intrusion detection, peer-to-peer data transfer, multi-hop relaying, packet forwarding in sensor networks, and many other[s]''}
\cite[p. 220]{GameTheoryCommunicationb00k2012}. In particular, a range of settings seems to be fruitfully modeled by means of coalition formation games \cite{Slikker2001}, which combine strategic and collaborative behavior, in that players choose what coalition to join, while  behaving in a `fully cooperative' manner within the chosen coalition. Such a modeling choice seems
grounded on the evidence that \textit{``recently, there has been a significant increase of interest in designing autonomic communication systems. Autonomic systems are networks
that are self-configuring, self-organizing, self-optimizing, and self-protecting. In such networks, the users should be able to learn and adapt to their environment (changes in
topology, technologies, service demands, application context, etc), thus providing much needed flexibility and functional scalability''} \cite{CoalitionalGameThoeryTutorial}. All
of this sounds extremely hard -if not impossible- to achieve without the absolute and constant cooperation of nodes with each other according to an agreed protocol that best fits the
needs and scope of the whole system. However, in order for choices to be rational (thereby allowing to discuss equilibria and related stability concerns), players have to be
rewarded depending on their choices. How to reward them in partition function-based coalition formation games is precisely (and almost exclusively) the object dealt with in the
present work, as the central role played by strategic equilibria in non-cooperative settings is replaced in cooperative ones with solutions or values, namely with
mappings specifying how to share the surplus of cooperation between players. Solutions of cooperative games thus address the same stability issue as equilibria of non-cooperative
games, since the idea is that for relevant generated surplus (i.e. given by, say, a supermodular lattice function) there are meaningful (i.e. fair, efficient, etc.) solutions
leading everybody to cooperate.

\subsection{Related work}
Insofar as TU (transferable utility) games are concerned, cooperative communication networking is frequently modeled by means of coalitional games (see \textsl{Class I} in
\cite[Section 7.2]{GameTheoryCommunicationb00k2012}), namely functions taking real values on the Boolean lattice \cite{Aigner97} of subsets or coalitions of players. This is the
traditional setting where TU games are mostly known, and the Shapley value is a fundamental solution \cite{Roth88}. However, recently attention has been placed also on more
complex games involving the geometric (indecomposable) lattice \cite{Aigner97} of partitions of nodes \cite{CoalitionalGameThoeryTutorial}. A partition (or coalition structure) is
a family of pairwise disjoint (nonempty) subsets of $N$, called blocks (or clusters, in the present framework), whose union is $N$. Many environments are characterized by
space-time node dynamics and technological means resulting in a clustered wireless network $G^t$ at each time $t$. That is, active nodes may be partitioned or clustered in
conformity with the given communication technology \cite{CoopSpecPhysComm2011,DistributedSpaceTimeProtocols2003,CoopSpecSens2007,EnergyEfficientIEEE2008}. For these settings, the
proposed approach relies mostly on distributed P2P communication systems where all (active) nodes behave in a collaborative manner, and seems to best fit multi-hop scenarios,
where it may also provide an additional perspective for identifying cluster heads
\cite{RoutingTechniques,ProceedingsShapleyValue,IEEEWirelessCommunications2005,TrustMilitary,GroCoca2007IEEE,RobustCooperativeRoutingProtocol,EnergyEfficientClusteringScheme2007,CooperativeDownloadingVehicularAdhocwirelessnetworks2005,JointPowerAllocationCooperative,DistributedRelaySelectionMultiuser,ComplexFieldNetworkCoding,HEED2004,NodeClusteringinWirelessSensorNetworks}. 
The foundation is perhaps best summarized by the subtitle of \cite{BookEgoisticallyCooperate}, i.e. \textit{``real egoistic behavior is to cooperate!''}. When the whole system
itself is very worthy to collaborating users, how to share such a worth is precisely the purpose of point-valued solutions (possibly in conjunction with set-valued ones such as
the core \cite{IEEEtransVeicTech2011,PartitionFormSpecSensIEEE2012}, see below). The dual perspective applies as well: if network maintainance is (computationally) demanding,
point-valued solutions also enable to fairly and efficiently share the corresponding costs. How a constant global cooperation in communication networking is quantifiable as a
partition function and why this may be meaningfully modeled via coalition formation are both comprehensively explained in
\cite{GameTheoryCommunicationb00k2012,CoalitionalGameThoeryTutorial}, hence these topics are not discussed here.
On the other hand, what emerges from the novel solutions proposed in the sequel is that the edges of the network may be looked at as
the true players in partition function-based TU games. Accordingly, these proposed solutions are mappings that share the surplus generated by partitions \textit{primarily} between
such edges. Of course, since edges do not gain from receiving an amount of TU, their shares are going to be further divided between the corresponding endvertices \cite{Borm++92}.
While providing a reward criterion that enforces trust via automated reciprocal control because of the involvement of two players for each share, this is also consistent with a
strictly game-theoretical perspective in view of the following combinatorial argument.

In TU cooperative game theory, partition functions are global games \cite{GilboaLehrer90GG}, meant to model cooperation over global issues such as environmental clean-up
and preservation. Specifically, every coalition structure or partition $P$ of players has an associated surplus or worth $f(P)$, to be interpreted as the level of satisfaction
common to all players attained when cooperation operates through $P$. Partitions of players or nodes are elements of an atomic lattice \cite{Aigner97} whose atoms are in fact the
$\binom{n}{2}$ unordered pairs $\{i,j\}\in N_2$ of nodes, namely the edges of the network. In the same way, singleton nodes or players $\{i\},i\in N$ are atoms in the
Boolean lattice where coalitional games take their values. The solutions proposed in the sequel are defined in terms of lattice functions with M\"obius inversion living only on
atoms \cite{Rossi2007HomoOeconomicus}, and this means mapping any given coalitional game into $n$ shares, one for each node, and any given global game into $\binom{n}{2}$ shares,
one for each edge. Thus for traditional set functions or coalitional games the novel definition simply rephrases the existing one, but for global games it leads to crucial
novelties. In this respect, communication networking is also sometimes modeled by means of a further type of TU cooperative games, known as `games in partition function form'
PFF (see \cite[p. 205]{GameTheoryCommunicationb00k2012} on \cite{ThrallLucas63}). These PFF games assign a worth to every embedded subset, namely to every coalition embedded into
a partition as a block, and might appear quite puzzling, especially in terms of their solutions (see \cite{GrabischFunaki,Myerson88} among others). A recent approach
\cite{Rossi2017GeometricLattice} shows that embedded subsets may be dealt with as elements of a lattice isomorphic \cite{Aigner97} to the partition lattice; specifically, PFF
games on $n$ players are combinatorially equivalent to global games on $n+1$ players. Hence the proposed solutions apply invariately to both global and PFF games, as explained
below insofar as possible, especially through an example fully devoted to the paralleling. These different TU games are formalized as lattice functions in Section II
hereafter, while also briefly detailing the well-known Shapley value of coalitional games as well as the main combinatorial aspects of global and PFF games. Next Section III
introduces the novel solution concept in terms of M\"obius inversion and atoms, showing how partition function-based TU cooperative games allow for two distinct Shapley values
because of the linear dependence \cite{Aigner97,Whitney35} characterizing the partition lattice, and with Subsection III.A devoted to symmetric games (i.e. functions
\cite{RosasSagan2006}), which seem possibly useful to model the surplus generated by cooperation in certain communication networking systems and whose solutions are determined
in a straightforward manner. Section IV shows how PFF games on $n$ players are isomorphic to global games on $n+1$ players as long as the lattice of embedded subsets is
taken to be geometric following \cite{Rossi2017GeometricLattice}, and the isomorphism is computationally detailed by means of an example. Section V translates the proposed
point-valued solutions in terms of the core \cite{Sandholm+2006,Shapley71}, which is the main set-valued solution concept, while showing that supermodularity is no longer
sufficient for its non-emptiness and by also briefly considering the case of additive \cite{FortunatoLongSurvey2010} or additively separable
\cite{GilboaLehrer90GG,GilboaLehrer91VI} partition functions. Section VI contains the conclusion.

\section{TU cooperative games as lattice functions}
TU cooperative games are functions taking real values on some lattice $(L^N,\wedge,\vee)$ grounded on player set $N$. Standard C (coalitional) games $v:2^N\rightarrow\mathbb R$ are
set functions defined on Boolean lattice $(2^N,\cap,\cup)$, where $2^N=\{A:A\subseteq N\}$ is the $2^n$-set of coalitions ordered by inclusion $\supseteq$. The meet $\wedge$ and
join $\vee$ of any two subsets $A,B\in 2^N$ respectively are intersection $A\cap B$ and union $A\cup B$. Othe other hand, global G games are partition functions
$f:\mathcal P^N\rightarrow\mathbb R$, i.e. defined on the geometric indecomposable \cite{Aigner97} lattice $(\mathcal P^N,\wedge,\vee)$ whose elements
$P=\{A_1,\ldots,A_{|P|}\},Q=\{B_1,\ldots,B_{|Q|}\}\in\mathcal P^N$ consist of blocks $A\in P,B\in Q$, namely nonempty and pairwise disjoint subsets $A,B\in 2^N$ whose union equals
$N$, hence $A\cap A'=\emptyset$ for $A,A'\in P$ while $A_1\cup\cdots\cup A_{|P|}=N$. Partitions $P,Q$ are ordered by coarsening $\geqslant$, i.e. $P\geqslant Q$ (or $P$ is coarser
than $Q$) if every block $B\in Q$ is included in some block $A\in P$, i.e. $A\supseteq B$. Meet $P\wedge Q$ and join $P\vee Q$ respectively are the coarsest partition finer than
both $P,Q$ and the finest partition coarser than both $P,Q$. Lattices $(2^N,\cap,\cup),(\mathcal P^N,\wedge,\vee)$ are atomic, since every element decomposes as the join of those
elements immediately above the bottom in the covering graph (or Hasse diagram) of the lattice, i.e the atoms \cite{Aigner97}. Among subsets, $\emptyset$ is the bottom and the $n$
singletons $\{i\},i\in N$ are the atoms.

\subsection{The Shapley value}
The Shapley value $\phi^{Sh}(v)$ is a fundamental solution of C games $v$ \cite{Roth88}. Geometrically, $\phi^{Sh}:\mathbb R^{2^n}\rightarrow\mathbb R^n$ is a mapping with
$\phi^{Sh}(v)=(\phi^{Sh}_1(v),\ldots,\phi^{Sh}_n(v))\in\mathbb R^n$ defined by
\begin{equation}
\phi^{Sh}_i(v)=\sum_{A\subseteq N\backslash i}\frac{v(A\cup i)-v(A)}{n\binom{n-1}{|A|}}=\sum_{A\subseteq N\backslash i}\frac{\mu^v(A\cup i)}{|A|+1}\text ,
\end{equation}
where $i\in N$ and $\mu^v:2^N\rightarrow\mathbb R$ is the M\"obius inversion of $v$, i.e. $\mu^v(A)=\sum_{B\subseteq A}(-1)^{|A\backslash B|}v(B)$  \cite{Aigner97,Rota64}. The
values taken by $\mu^v$ are also sometimes referred to as `Harsanyi dividends' \cite{Roth88}. The former expression obtains by
placing the uniform (probability) distribution over the $n!$ maximal chains in lattice $(2^N,\cap,\cup)$, and considering marginal contributions $v(A\cup i)-v(A)$
of players $i\in N$ to coalitions $A\subseteq N\backslash i$. A maximal chain is a $n+1$-set $\{A_0,A_1,\ldots,A_n\}\subset2^N$ such that $|A_k|=k,0\leq k\leq n$ and
$A_{l+1}\supset A_l,0\leq l<n$, where $\supset$ is proper inclusion, hence $A_{l+1}$ \textit{covers} $A_l$ \cite{Aigner97}. In this view, $\phi^{Sh}_i(v)$ is the expectation
of the marginal contribution of $i$ to a random coalition $A\subseteq N\backslash i$, as detailed in \cite{Weber88} in terms of probabilistic and random-order values.
The latter expression regards $\mu^v(A)$ as the net added worth (possibly $<0$) of cooperation within coalition $A$ with respect to all its proper subcoalitions $B$. That is,
$\mu^v(A)=v(A)-\sum_{B\subset A}\mu^v(B)$. This net added worth is equally shared between coalition members $i\in A$ according to the following well-known axiomatic
characterization of $\phi^{Sh}$ ($v,v'$ are C games, $\alpha,\beta\in\mathbb R$, $i,j\in N$).\\
\textit{L (linearity):} $\phi(\alpha v+\beta v')=\alpha\phi(v)+\beta\phi(v')$.\\
\textit{S (symmetry):} if $v(A\cup i)=v(A\cup j)$ for all $A\subseteq N\backslash\{i,j\}$, then $\phi_i(v)=\phi_j(v)$.\\
\textit{D (dummy):} if $v(A\cup i)=v(A)+v(\{i\})$ for all $A\subseteq N\backslash i$, then $\phi_i(v)=v(\{i\})$.\\
\textit{E (efficiency):} $\sum_{i\in N}\phi_i(v)=v(N)$.\\
Indexing axes by coalitions $A\in 2^N$, C games $v\in\mathbb R^{2^n}$ are points in a vector space. A main basis of this space is $\{\zeta_A:A\in2^N\}$, where
$\zeta_A(B)=\left\{\begin{array}{c}1\text{ if }B\supseteq A\\0\text{ if }B\not\supseteq A\end{array}\right.$ is the zeta function $\zeta_A$ \cite{Aigner97,Rota64} (also termed
unanimity game $u_A$ \cite{Roth88}). Any $v$ is the following linear combination of basis elements:
$v(\cdot)=\sum_{A\in 2^N}\zeta_A(\cdot)\mu^v(A)$ or $v(B)=\sum_{A\subseteq B}\mu^v(A)$. If $\phi$ satisfies $L$, then
$\phi(v)=\sum_{A\in 2^N}\mu^v(A)\phi(\zeta_A)$. The focus can thus be placed on the implications of $S,D,E$ for any $\zeta_A$. Now, $D$ entails
$\phi_j(\zeta_A)=0$ for all $j\in A^c=N\backslash A$, while $\phi_i(\zeta_A)=\phi_{i'}(\zeta_A)$ for all $i,i'\in A$ in view of $S$. Finally, $E$ requires
$\phi_i(\zeta_A)=\frac{1}{|A|}$ for all $i\in A$. Observe that the fixed points of the Shapley value mapping are those C games $v$ such that $v(A)=\sum_{i\in A}v(\{i\})$ for all
coalitions $A\in 2^N$. Such set functions $v$ are valuations \cite{Aigner97} of Boolean lattice $(2^N,\cap,\cup)$, satisfying $v(A\cup B)+v(A\cap B)=v(A)+v(B)$ for all
$A,B\in 2^N$. Their M\"obius inversion satisfies $\mu^v(A)=0$ for all $A\in 2^N,|A|>1$. Furthermore,
$\mu^{\zeta_A}(B)=\left\{\begin{array}{c}1\text{ if }B=A\\0\text{ otherwise}\end{array}\right.$.

\subsection{Global and PFF games}
The bottom partition is $P_{\bot}=\{\{1\},\ldots,\{n\}\}$, with rank $r(P_{\bot})=n-|P_{\bot}|=0$, while the $\binom{n}{2}$ atoms have rank 1 and consist each of $n-1$ blocks,
i.e. $n-2$ singletons and one (unordered) pair. Denote by $[ij]\in\mathcal P^N$ the atom whose non-singleton block is pair $\{i,j\}\in N_2$. A main difference between
$(2^N,\cap,\cup)$ and $(\mathcal P^N,\wedge,\vee)$ relies in the join-decomposition \cite{Aigner97} of their elements, in that every $A\in 2^N$ decomposes uniquely as
$A=\cup_{i\in A}\{i\}$, while most partitions admit several decompositions as a join of atoms. In fact, $r(A)=|A|$ is precisely the number of atoms involved in the
join-decomposition of subset $A$, while $r(P)=n-|P|$ is the minimum number of atoms involved in the join-decomposition of partition $P$. This is immediately seen for the top
partition which consists of a single block, i.e. $P^{\top}=\{N\}$, in that $P^{\top}=[ij]_1\vee\cdots\vee[ij]_{n-1}$ for any $n-1$ atoms satisfying 
$|\{ij\}_k\cap\{ij\}_m|=\left\{\begin{array}{c}1\text{ if }m=k+1\\0\text{ otherwise}\end{array}\right.$, $1\leq k<m<n$, whereas clearly $P^{\top}\geqslant[ij]$ for all the
$\binom{n}{2}$ atoms, entailing that $P^{\top}=[ij]_1\vee\cdots\vee[ij]_{\binom{n}{2}}$ too. Define the maximum number of atoms involved in the join-decomposition of partitions to
be the size $s:\mathcal P^N\rightarrow\mathbb Z_+$ of these latter. That is to say, $s(P)=|\{[ij]:P\geqslant[ij]\}|$. Evidently, $s(P)\geq r(P)$, with equality if and only if $P$
has no blocks larger than pairs.

As outlined above, G (global) games are partition functions $f:\mathcal P^N\rightarrow\mathbb R$, with $f(P)$ quantifying the surplus generated by cooperation when players are
operating through coalition structure $P$. However, as detailed in the sequel, single players $i\in N$ factually may not provide any marginal contribution in G games. In terms
of communication networks, they are isolated vertices, hence either inactive (at some time $t$ under concern) or else not able to communicate with anybody because of their
spatial position. Conversely, atoms essentially coincide with pairs of players, i.e. hops or edges, and they do contribute to the functioning of the network as vehicles for information exchange. In other terms, in order for an edge to be worthy to the collective, \textit{both} endvertices must collaborate.

The zeta function $\zeta_P(Q)=\left\{\begin{array}{c}1\text{ if }Q\geqslant P\\0\text{ otherwise}\end{array}\right.$ is the analog basis element as $\zeta_A,A\in 2^N$ and again
M\"obius inversion $\mu^f(P)=f(P)-\sum_{Q<P}\mu^f(Q)$ provides the coefficients of the linear combination of these basis elements, where $<$ is proper coarsening and
$\mu^{\zeta_P}(Q)=\left\{\begin{array}{c}1\text{ if }Q=P\\0\text{ otherwise}\end{array}\right.$. That is to say, $f(\cdot)=\sum_{P\in\mathcal P^N}\mu^f(P)\zeta_P(\cdot)$ or
$f(Q)=\sum_{P\leqslant Q}\mu^f(P)$. Concerning marginal contributions of atoms $[ij]$, if $P\not\geqslant[ij]$, then $P\vee[ij]$ covers $P$, usually denoted \cite{Aigner97} by
$P\vee[ij]\gtrdot P$, and $P\vee[ij]$ obtains by merging those two blocks $A,A'\in P$ such that $A\cap\{i,j\}=\{i\},A'\cap\{i,j\}=\{j\}$. Maximal chains of partitions are
collections $\{P_0,P_1,\ldots,P_{n-1}\}$ where $P_k\gtrdot P_{k-1}$, hence $r(P_k)=k,0\leq k<n$, with $P_0=P_{\bot},P_{n-1}=P^{\top}$. There are $\frac{n!(n-1)!}{2^{n-1}}$
(distinct) maximal chains of partitions.

PFF games may be denoted by $h$ and are more complex lattice functions, taking their values on embedded subsets, namely (ordered) pairs $(A,P)\in2^N\times\mathcal P^N$ such that
$A\in P$, i.e $A$ is a block of $P$. Although $\emptyset$ is not a block of any partition, still M\"obius inversion $\mu^h$ needs a bottom element, and thus $(\emptyset,P_{\bot})$
is taken to be the bottom embedded subset \cite{Grabisch2010DAM}. Combinatorial congruhence then requires $(\emptyset,P)$ to be an embedded subset for
\textit{all} $P\in\mathcal P^N$, otherwise the resulting lattice is non-atomic. In this way, the family $\mathcal E^N\subset2^N\times\mathcal P^N$ of embedded
subsets $(A,P)$ such that either $A\in P$, or else $A=\emptyset$, is a geometric lattice isomorphic \cite{Aigner97} to $\mathcal P^{N_+}$, where $N_+=\{1,\ldots,n,n+1\}$. The
$\binom{n+1}{2}=n+\binom{n}{2}$ atoms of $\mathcal E^N$ are the $n$ pairs $(i,P_{\bot})$ together with the $\binom{n}{2}$ pairs $(\emptyset,[ij])$. Denote by
$(\mathcal E^N,\sqcap,\sqcup)$ this lattice, with order relation $\sqsupseteq$; it is comprehensively detailed in \cite{Rossi2017GeometricLattice}. Given the isomorphism, there
are $\frac{(n+1)!n!}{2^n}$ maximal chains in $\mathcal E^N$, along which atoms provide marginal contributions like in C and G games above.

\section{Solutions}
Denote by $(L^N,\wedge,\vee)$ a lattice grounded on player set $N$, namely $L^N\in\{2^N,\mathcal P^N,\mathcal E^N\}$, with elements $x,y,z,\ldots\in L$, order relation
$\geqslant$ and bottom element $x_{\bot}$. Let $L^N_{\mathcal A}$ be the set of atoms of $L^N$, while C, G and PFF TU cooperative games may now be dealt with at once as
lattice functions $f:L^N\rightarrow\mathbb R$. Recall that $f:L^N\rightarrow\mathbb R$, with M\"obius inversion $\mu^f:L^N\rightarrow\mathbb R$, is totally positive if
$\mu^f(x)\geq 0$ for all $x\in L^N$, while if $f(x\wedge y)+f(x\vee y)\geq f(x)+f(y)$ for all $x,y\in L^N$, then $f$ is supermodular. Total positivity is a sufficient (but not
necessary) condition for supermodularity. Valuations of $(L^N,\wedge,\vee)$ are those $f$ satisfying supermodularity with equality, i.e. $f(x\wedge y)+f(x\vee y)=f(x)+f(y)$ for
all $x,y\in L^N$. As already explained, for coalitional games $v$, a solution is a mapping $\phi$ associating with $v$ a further coalitional game $\phi(v)$ which is a valuation of
subset lattice $(2^N,\cap,\cup)$. Hence $\phi(v)$ is an \textit{inessential} game, as every player is a dummy: $\phi(v)(A)=\sum_{i\in A}\phi_i(v)$. That is, $\phi(v)$ is a coalitional game
formalizing a situation where there is no surplus generated by cooperation. A standard and most natural assumption is  $v(\emptyset)=0=\phi(v)(\emptyset)$. Then, given general result \cite[Theorem 4.63, p. 190]{Aigner97} for valuations of distributive lattices, solutions $\phi(v)$ of coalitional games $v$ may be equivalently defined to be coalitional
games whose M\"obius inversion $\mu^{\phi(v)}$ lives only on atoms: $\mu^{\phi(v)}(A)=0$ for all $A\in 2^N$ such that $|A|\neq 1$.

Both $\mathcal P^N$ and $\mathcal E^N$ are geometric indecomposable \cite[p. 61]{Aigner97}, and thus valuations of these lattices are constant functions
\cite[Exercise IV.4.12, p. 195]{Aigner97}. Accordingly, solutions of G and PFF games cannot be defined in terms of valuations of $\mathcal P^N$ and $\mathcal E^N$, respectively.
In fact, solutions of G games were defined \cite{GilboaLehrer90GG} in terms of valuations of subset lattice $(2^N,\cap,\cup)$, with the implication that only the worth
$f(P^A_{\bot})$ of those $2^n-n$ partitions $P^A_{\bot}=\{A,\{i_1\},\ldots,\{i_{n-|A|}\}\}$ with at most one non-singleton block is taken into account, where
$A^c=\{i_1,\ldots,i_{n-|A|}\}$. Thus $\mathcal B_n-(2^n-n)$ values taken by $f$ are disregarded, $\mathcal B_n$ being the (Bell) number of partitions of a
$n$-set \cite{Aigner97}. The same argument applies to existing solutions of PFF games \cite{GrabischFunaki,Myerson88}. For these reseasons, in the remainder of this work solutions
of TU cooperative games are conceived in terms of M\"obius inversion of lattice functions \cite{Rossi2007HomoOeconomicus} as follows.
\begin{definition}
Solutions of cooperative games $f:L^N\rightarrow\mathbb R$ are mappings $\phi:f\rightarrow\phi(f)$ associating with $f$ an analog game
$\phi(f):L^N\rightarrow\mathbb R$ whose M\"obius inversion $\mu^{\phi(f)}:L^N\rightarrow\mathbb R$ lives only on atoms, i.e. $\mu^{\phi(f)}(x)=0$ for all
$x\in L^N\backslash L^N_{\mathcal A}$.
\end{definition}
Thus a solution of a game is a lattice function taking values on the same lattice where the game itself takes its values. In this way a solution is a game as well, and in
particular one where cooperation no longer generates any surplus. This seems the only combinatorially consistent way to include G and PFF games within the standard framework of C
games. Let $a\in L^N_{\mathcal A}$ be the generic atom of $L^N$ and $\phi_a(f)=\phi(f)(a)$. Then,
\begin{equation}
\phi(f)(x)=\sum_{a\leqslant x}\phi_a(f)\text{ for all }x\in L^N\text .
\end{equation}

\begin{example}
Consider a simplest G game $f:\mathcal P^N\rightarrow\mathbb R$ defined by $f=\binom{n}{2}\zeta_{P^{\top}}$. That is,
$f(P)=\left\{\begin{array}{c}\binom{n}{2}\text{ if }P=P^{\top}\\0\text{ if }P<P^{\top}\end{array}\right.$. The size $s=\phi(f)$ defined in Section II then seems the most
appropriate solution, as $s(P_{\bot})=0$ while on atoms $s([ij])=1=\phi_{[ij]}(f)$, and its M\"obius inversion satisfies
$\mu^s(Q)=\left\{\begin{array}{c}1\text{ if }s(Q)=1\\0\text{ if }s(Q)\neq 1\end{array}\right.$. Thus $\phi_{[ij]}(f)=\mu^s([ij])$ and
$$s(P)=\sum_{Q\leqslant P}\mu^s(Q)=\sum_{[ij]\leqslant P}\mu^s([ij])=\sum_{A\in P}\binom{|A|}{2}\text .$$
\end{example}

Following the above traditional axiomatic characterization of the Shapley value \cite{Roth88}, firstly consider L (linearity).
\begin{definition}
A solution $\phi$ is linear if $\phi(\alpha f)=\alpha\phi(f)$ for $\alpha\in\mathbb R$, and $\phi(f+f')=\phi(f)+\phi(f')$ for $f,f':L^N\rightarrow\mathbb R$.
\end{definition}
Since $\{\zeta_x:x\in L^N\}$ is a basis of the so-called \cite{Aigner97} free vector space $\mathbb R^{|L^N|}$ of lattice functions $f:L^N\rightarrow\mathbb R$, with
coefficients given by M\"obius inversion, i.e. $f(\cdot)=\sum_{x\in L^N}\mu^f(x)\zeta_x(\cdot)$ or $f(y)=\sum_{x\leqslant y}\mu^f(x)$, a solution satisfying L has form
\begin{equation}
\phi(f)=\sum_{x\in L^N}\mu^f(x)\phi(\zeta_x)\text .
\end{equation}
Such solutions are univocally defined by specifying how to distribute the unit of TU given by `zeta games'
$\zeta_x,x\in L^N$. 
\begin{definition}
Denoting by $x^{\top}$ the top element of $L^N$, a solution $\phi$ is efficient if $\sum_{a\in L^N_{\mathcal A}}\phi_a(f)=f(x^{\top})$.
\end{definition}
E (efficiency) was conceived \cite{Roth88} to deal with monotone (and superadditive \cite{Shapley53}) C games, in which case it seems a most natural assumption. But for G and PFF
games it requires some caution. In fact, a lattice function $f$ is monotone if for any $x,y\in L^N$ such that $x\geqslant y$, inequality $f(x)\geq f(y)$ holds, entailing
$f(x^{\top})\geq f(x)$ for all $x\in L^N$ (while superadditivity is neither straightforwardly translated nor interesting for G and PFF games). When modeling the surplus generated
by cooperation in clustered (multi-hop mobile) wireless networks as a G or PFF game \cite{GameTheoryCommunicationb00k2012,CoalitionalGameThoeryTutorial}, monotonicity would
basically mean that by putting all nodes into a unique grand cluster the surplus of cooperation attains its maximum. Evidently, this is not the case, as the
network is clustered (with an associated computational cost) precisely because partitioning the nodes enables for a better communication in view of the available technological
infrastructure. One way to deal with this, while maintaining E as a fundamental axiom for characterizing solutions, is by letting M\"obius inversion $\mu^f$ take value 0
on all partitions \textit{non-finer} than that identified via the given global clustering algorithm
\cite{IEEEWirelessCommunications2005,BookEgoisticallyCooperate,EnergyEfficientClusteringScheme2007,HEED2004,NodeClusteringinWirelessSensorNetworks,EnergyEfficientIEEE2008}. 
In terms of lattice functions, this is not much different from the above definition of solutions, as in both cases the basic modeling tool is M\"obius inversion: taking its
value to be identically 0 on a certain suitable region of the lattice formalizes the fact that some issues are autonomously addressed either by the agreed sharing criterion
or else by the given communication technology. Formally, if $P_t^*$ is the node partition defined by the chosen clustering algorithm at a generic time $t$, then
$\mu^f(Q)=0$ for all $Q\not\leqslant P^*_t$ yields
\begin{equation}
f(P^{\top})=\sum_{Q\in\mathcal P^N}\mu^f(Q)=\sum_{Q\leqslant P^*_t}\mu^f(Q)=f(P^*_t)\text .
\end{equation}

While L provides expression (3), E additionally entails that $\sum_{a\in L^N_{\mathcal A}}\phi_a(\zeta_x)=1$ for all basis elements or zeta games $\zeta_x,x\in L^N$. Concerning
D (dummy), G and PFF games are crucially different from C games in view of the linear dependence characterizing geometric lattices \cite{Aigner97,Whitney35}. In fact, there are
no dummy atoms in partition function-based games. To see this, consider any basis element $\zeta_P$ of G games and any atom $[ij]\in\mathcal P^N_{\mathcal A}$
($\mathcal P^N_{\mathcal A}$ being the $\binom{n}{2}$-set of atoms of $\mathcal P^N$). If $[ij]\leqslant P$, then\\
$\zeta_P(Q\vee[ij])-\zeta_P(Q)=\left\{\begin{array}{c}1\text{ if }Q\not\geqslant[ij]\text{ and }P=Q\vee[ij]\\0\text{ otherwise}\end{array}\right.$,\\
while if $[ij]\not\leqslant P$, then\\
$\zeta_P(Q\vee[ij])-\zeta_P(Q)=\left\{\begin{array}{c}1\text{ if }Q\not\geqslant P\text{ and }Q\vee[ij]>P\\0\text{ otherwise}\end{array}\right.$.\\
Observe that this also applies to those $\binom{n}{2}$ zeta games $\zeta_{[ij]}$ grounded on atoms $[ij]\in\mathcal P^N_{\mathcal A}$, as detailed hereafter.
\begin{example}
Let $N=\{1,2,3\}$, hence there are three atoms $[12],[13],[23]$ and five partitions, i.e. the bottom $P_{\bot}=1|2|3$ (with vertical bar $|$ separating blocks), the three atoms
and $P^{\top}=[12]\vee[13]=[12]\vee[23]=[13]\vee[23]=[12]\vee[13]\vee[23]$, namely the top. For $\zeta_{[12]}$, of course $\zeta_{[12]}([12])=1$. However,
$\zeta_{[12]}([13]\vee[23])=\zeta_{[12]}(P^{\top})=1$. This means that even if G game $\zeta_{[12]}$ requires atom $[12]$ to cooperate in order to generate the unit of TU, still
such a unit also obtains through the cooperation of `only' the other two atoms, namely $[13]$ and $[23]$, because if these latter cooperate then $[12]$ `must' cooperate too. As
detailed in the sequel, this implies that if the Shapley value of G games is translated according to the former equality in expression (1), then the unit of TU generated by zeta
games has to be distributed over \textit{all} atoms. Conversely, if the latter equality in expression (1) is employed, then the resulting solution is very different.
\end{example}

Maintaining the axiomatic characterization of the Shapley value outlined in Section II, in order to formalize S (symmetry) for G and PFF games recall
that the class $c^P$ (or type) of partitions $P\in\mathcal P^N$ \cite{Aigner97,Rota64,Stanley2012EnuCom} is $c^P=(c^P_1,\ldots,c^P_n)\in\mathbb Z_+^n$, where
$c^P_k=|\{A:k=|A|,A\in P\}|$ is the number of $k$-cardinal blocks of $P,1\leq k\leq n$. Analogously, the class of $(A,P)\in\mathcal E^N$ is
$c^{A,P}=(c^{A,P}_0,c^{A,P}_1,\ldots,c^{A,P}_n)\in\mathbb Z^{n+1}_+$, as embedded subset $A$ may have cardinality $0\leq|A|=c^{A,P}_0\leq n$
\cite{Rossi2017GeometricLattice}. Hence $x\in L^N$ has class $c^x$, for $L^N\in\{\mathcal P^N,\mathcal E^N\}$. Also, $s(x)=|\{a:L^N_{\mathcal A}\ni a\leqslant x\}|$ is
the size of lattice elements $x$. The size of $(A,P)\in\mathcal E^N$ thus is $s(A,P)=|A|+s(P)$.
\begin{definition}
A solution $\phi$ satisfies S if $\phi_a(f)=\phi_{a'}(f)$ whenever any two atoms $a,a'\in L^N_{\mathcal A}$ allow for a bijection between $\{x:L^N\ni x\not\geqslant a\}$ and
$\{y:L^N\ni y\not\geqslant a'\}$ satisfying (i) $f(x\vee a)=f(y\vee a')$ and (ii) $c^x=c^y$.
\end{definition}
Since D cannot be employed to characterize solutions of G and PFF games in view of Example 5, these remaining axioms L, E and S do not yield uniqueness, but conversely define a
whole class of solutions. In fact, given L and E, there is a continuum of alternative manners to also satisfy S, ranging from the following two extreme cases:
(a) $\phi_a(\zeta_x)=\left\{\begin{array}{c}s(x)^{-1}\text{ if }a\leqslant x\\0\text{ if }a\not\leqslant x\end{array}\right.$, and
(b) $\phi_a(\zeta_x)=|L^N_{\mathcal A}|^{-1}$ for \textit{all} atoms $a\in L^N_{\mathcal A}$ (and any $\zeta_x$). Within this broad class of solutions satisfying L, E and S, a
useful discriminant is the following FP \textit{fixed-point} condition (which is a variation of the D axiom applying to C games).
\begin{definition}
A solution $\phi$ satisfies FP if $\mu^f(x)=0$ for all $x\in L^N\backslash L^N_{\mathcal A}$ entails $\phi(f)=f$.
\end{definition}
Thus $\phi$ satisfies FP when it maps those games $f$ whose M\"obius inversion $\mu^f$ already lives only on atoms into themselves, i.e.
\begin{equation}
\phi_{a'}(\zeta_a)=\left\{\begin{array}{c}1\text{ if }a'=a\\0\text{ if }a'\neq a\end{array}\right.\text{ for all }a,a'\in L^N_{\mathcal A}\text .
\end{equation}
In comparison with Example 5, this expression states that even if in G and PFF games every single atom $a'$ may `swing'\footnote{See \cite{Roth88} on the Banzhaf value of C games.}
in the zeta game $\zeta_a$ grounded on a fixed atom $a$, still FP requires $a$ to exclusively get the whole unit of TU generated by $\zeta_a$.

The remainder of this work is mostly concerned with the following two solutions $\phi^{CU},\phi^{SU}$ satisfying L, E and S.
\begin{definition}
The chain-uniform CU solution $\phi^{CU}$ is
\begin{equation}
\phi_a^{CU}(f)=\sum_{x\not\geqslant a}\frac{\kappa_x^a}{\kappa}\left(\frac{f(x\vee a)-f(x)}{s(x\vee a)-s(x)}\right)\text ,
\end{equation}
where $\kappa_x^a/\kappa$ is the ratio of the number $\kappa_x^a$ of maximal chains (in $L^N$) meeting both $x$ and $x\vee a$ to the total
number $\kappa$ of maximal chains, thus $\sum_{x\not\geqslant a}\frac{\kappa_x^a}{\kappa}=1$.\smallskip\\
The size-uniform SU solution $\phi^{SU}$ is
\begin{equation}
\phi_a^f(SU)=\sum_{x\geqslant a}\frac{\mu^f(x)}{s(x)}\text .
\end{equation}
\end{definition}

For C games $v$, the CU and SU solutions coincide with the Shapley value, i.e. $\phi^{CU}(v)=\phi^{Sh}(v)=\phi^{SU}(v)$, where
$\frac{1}{n\binom{n-1}{|A|}}=\frac{|A|!(n-|A|-1)!}{n!}$ is the ratio of the number of maximal chains meeting both $A\subseteq N\backslash i$ and $A\cup i$ to the total number $n!$
of maximal chains, while $|A\cup i|-|A|=1=s(A\cup i)-s(A)$ is the size change. Conversely, for G and PFF games these two solutions are very different, and in particular $\phi^{SU}$
satisfies FP while $\phi^{CU}$ does not. Explicitely, for any zeta game $\zeta_y,y>x_{\bot}$,
\begin{equation}
\phi_a^{SU}(\zeta_y)=\left\{\begin{array}{c} \frac{1}{s(y)}\text{ if }a\leqslant y\\0\text{ otherwise} \end{array}\right.\text{ for all }a\in L^N_{\mathcal A}\text .
\end{equation}
Hence when $y=a$ expression (5) applies. On the other hand, $$f(x\vee a)-f(x)=\sum_{x\not\geqslant y\leqslant x\vee a}\mu^f(y)\text{ yields}$$
\begin{eqnarray*}
\phi_a^{CU}(f)&=&\sum_{x\not\geqslant a}\frac{\kappa_x^a}{\kappa}\left(\frac{\sum_{x\not\geqslant y\leqslant x\vee a}\mu^f(y)}{s(x\vee a)-s(x)}\right)\\
&=&\sum_{y\in L^N}\mu^f(y)\left[\sum_{\underset{x\not\geqslant y\leqslant x\vee a}{x\not\geqslant a}}\frac{\kappa_x^a}{\kappa\big[s(x\vee a)-s(x)\big]}\right]\text .
\end{eqnarray*}
Thus for any zeta game $\zeta_y,y>x_{\bot}$, 
\begin{equation*}
\phi_a^{CU}(\zeta_y)=\sum_{\underset{x\not\geqslant y\leqslant x\vee a}{x\not\geqslant a}}\frac{\kappa_x^a}{\kappa\big[s(x\vee a)-s(x)\big]}
\text{ for all }a\in L_{\mathcal A}\text ,
\end{equation*}
entailing
% \begin{eqnarray*}
% \sum_{a\in L_{\mathcal A}}\phi_a^f(CU)&=&
% \sum_{y\in L}\mu^f(y)\left(\sum_{a\in L_{\mathcal A}}\sum_{\underset{x\not\geqslant y\leqslant x\vee a}{x\not\geqslant a}}\frac{\kappa_x^a}{\kappa_L(s(x\vee a)-s(x))}\right)=\\
% &=&\sum_{y\in L}\mu^f(y)\left(\sum_{\underset{x\vee a\geqslant y\text{ for some atom }a\not\leqslant x}{x\not\geqslant y}}\frac{\kappa_x^a}{\kappa_L}\right)=
% \sum_{y\in L}\mu^f(y)=f(x^{\top})\text .
% \end{eqnarray*}
% The CU solution is thus also efficient, like the SU one. Yet, it does not satisfy the fixed-point condition, in that
\begin{eqnarray*}
\phi_a^{CU}(\zeta_a)&=&\sum_{x\not\geqslant a}\frac{\kappa_x^a}{\kappa\big[s(x\vee a)-s(x)\big]}\leq\sum_{x\not\geqslant a}\frac{\kappa_x^a}{\kappa}=1\text ,\\
%\text{ for all }a\in L_{\mathcal A}\\
\phi_{a'}^{CU}(\zeta_a)&=&\sum_{\underset{x\vee a=x\vee a'}{x\not\geqslant a,a'}}\frac{\kappa_x^a}{\kappa\big[s(x\vee a)-s(x)\big]}\geq 0\text .
%\text{ for all }a'\in L_{\mathcal A},a'\neq a\text .
\end{eqnarray*}
For subset lattice $L^N=2^N$, these two inequalities are satisfied as equalities, in that $\{x:x\not\geqslant a,a',x\vee a=x\vee a'\}=\emptyset$ for any two distinct atoms
$a,a'$ as well as $s(x\vee a)-s(x)=1$, thus the SU and CU solutions coincide on these basis elements $\zeta_{\{i\}},\{i\}\in 2^N$ of C games. Conversely, for G and PFF games the above
inequalities are strict, i.e. $s(x\vee a)-s(x)>1$ for most $x\not\geqslant a$ and $\{x:x\not\geqslant a,a',x\vee a=x\vee a'\}\neq\emptyset$ for any two distinct atoms $a,a'$.
Hence the SU and CU solutions are different and the latter does not satisfy FP: $\phi_a^{CU}(\zeta_a)<1$ and $\phi_{a'}^{CU}(\zeta_a)>0$ for all $a,a'\in L^N_{\mathcal A}$.
In practice, the only fixed points of the CU solution are also fixed points of the SU solution and take the form of linear functions $f=\alpha s,\alpha>0$ of the size $s$. That is,
$\phi_a^{CU}(\alpha s)=$
\begin{equation*}
=\sum_{x\not\geqslant a}\frac{\kappa_x^a}{\kappa}\left(\frac{\alpha\big[s(x\vee a)-s(x)\big]}{s(x\vee a)-s(x)}\right)=
\alpha\sum_{x\not\geqslant a}\frac{\kappa_x^a}{\kappa}=\alpha
\end{equation*}
for all $a\in L^N_{\mathcal A}$.

\subsection{Symmetric games}
An important class of games that may be useful for modeling certain communication networking systems consists of symmetric ones. In fact, \textit{`partitions are of central
importance in the study of symmetric functions, a class of functions that pervades mathematics in general'} \cite[p. 39]{Knuth2005} (see also \cite[Ch. 5]{RotaWay2009},
\cite{RosasSagan2006}, \cite[Ch. 7, Vol. 2]{Stanley2012EnuCom}). As detailed hereafter, for G and PFF symmetric games the CU and SU solutions coincide.
\begin{definition}
C games $v$, G games $f$ and PFF games $h$ are symmetric if
\begin{itemize}
\item[$\bullet$] $|A|=|B|$ entails $v(A)=v(B)$,
\item[$\bullet$] $c^P=c^Q$ entails $f(P)=f(Q)$,
\item[$\bullet$] $c^{A,P}=c^{B,Q}$ entails $h(A,P)=h(B,Q)$.
\end{itemize}
\end{definition}
Such $v,f$ and $h$ are indeed invariant under the action of the symmetric group $\mathcal S_n$ whose elements are the $n!$ permutations
$\pi:\{1,\ldots ,n\}\rightarrow\{1,\ldots ,n\}$ of the indices or node identifiers $i\in N$. In particular, for every (number) partition
$(\lambda_1,\ldots,\lambda_n)\in\mathbb Z^n_+$ of (integer) $n$ (i.e., $\sum_{1\leq k\leq n}\lambda_k=n$), the number of distinct (set) partitions $P$ of $N$ with class
$c^P_k=\lambda_k$, $1\leq k\leq n$ is $n!\left(\prod_{1\leq k\leq n}k!^{c^P_k}c^P_k!\right)^{-1}$ \cite[Vol. 1, p. 319]{Stanley2012EnuCom}.

In this view, if $f:L^N\rightarrow\mathbb R$ is a symmetric lattice function (where $L^N\in\{\mathcal P^N,\mathcal E^N\}$), then
\begin{equation*}
\phi^{SU}_a(f)=\phi^{SU}_{a'}(f)\text{ or }\sum_{x\geqslant a}\frac{\mu^f(x)}{s(x)}=\sum_{y\geqslant a'}\frac{\mu^f(y)}{s(y)}
\end{equation*}
for any two atoms $a,a'\in L^N_{\mathcal A}$. Accordingly, in view of E,
$$\sum_{a\in L^N_{\mathcal A}}\phi^{SU}_a(f)=|L^N_{\mathcal A}|\phi^{SU}_a(f)=f(x^{\top})\Rightarrow\phi_a^{SU}(f)=\frac{f(x^{\top})}{|L^N_{\mathcal A}|}\text .$$

Analogously, a symmetric $f$ yields $\phi^{CU}_a(f)=\phi^{CU}_{a'}(f)$ as
$$\sum_{x\not\geqslant a}\frac{\kappa_x^a}{\kappa}\left[\frac{f(x\vee a)-f(x)}{s(x\vee a)-s(x)}\right]=$$
$$=\sum_{y\geqslant a'}\frac{\kappa_y^{a'}}{\kappa}\left[\frac{f(y\vee a')-f(y)}{s(y\vee a')-s(y)}\right]\text{for all }a,a'\in L_{\mathcal A}\text{, hence}$$
$$\sum_{a\in L^N_{\mathcal A}}\phi^{CU}_a(f)=|L^N_{\mathcal A}|\phi^{CU}_a(f)=f(x^{\top})\Rightarrow$$
$\Rightarrow\phi_a^{CU}(f)=f(x^{\top})/|L_{\mathcal A}|$. Basic symmetric partition functions or G games (or PFF games, given the isomorphism exemplified in section IV below)
are the size $s(P)=\sum_{A\in P}\binom{|A|}{2}$ and the rank $r(P)=n-|P|=\sum_{A\in P}(|A|-1)$. Therefore,
\begin{equation*}
\phi_a^{SU}(r)=\phi_a^{CU}(r)=\frac{r(x^{\top})}{|L^N_{\mathcal A}|}=\phi_a^{CU}(\zeta_a)\leq 1=\phi_a^{SU}(\zeta_a)
\end{equation*}
for all atoms $a$, with strict inequality if $L^N\in\{\mathcal P^N,\mathcal E^N\}$. More generally, any $f$ which is itself a function of the rank or a function of the size (see
Example 2) is of course symmetric, and thus satisfies $\phi_a^{CU}(f)=f(x^{\top})/|L^N_{\mathcal A}|=\phi_a^{SU}(f)$.

\section{Isomorphism between G and PFF games}
This section details the computations for the SU and CU solutions of G and PFF games, while also providing a simple example of G games on player set $N=\{1,2,3\}$ together with the isomorphic PFF games on player set $N=\{1,2\}$.

For G games $f:\mathcal P^N\rightarrow\mathbb R$, solutions are $\binom{n}{2}$-vectors $\phi(f)=\{\phi_{[ij]}(f):\{i,j\}\in N_2\}\in\mathbb R^{\binom{n}{2}}$. Regarded as lattice
functions themselves, these solutions have M\"obius inversion $\mu^{\phi(f)}([ij])=\phi_{[ij]}(f)$ living only on atoms $[ij]$, and thus satisfy
$\phi(f)(P)=\sum_{[ij]\leqslant P}\phi_{[ij]}(f)$ for all $P\in\mathcal P^N$.

There are $\kappa=\binom{n}{2}\binom{n-1}{2}\binom{n-2}{2}\cdots\binom{2}{2}=\frac{n!(n-1)!}{2^{n-1}}$ maximal chains of partitions. The number of such maximal chains meeting any
$P$, with class $c^P$, is
\begin{equation*}
\left[\prod_{1\leq k\leq n}\left(\frac{k!(k-1)!}{2^{k-1}}\right)^{c^P_k}\right]\frac{|P|!(|P|-1)!}{2^{|P|-1}}\text ,
\end{equation*}
where the product on the left is the number of maximal chains in $[P_{\bot},P]$, this latter segment or interval \cite{Rota64} being isomorphic to
$\times_{A\in P}\mathcal P^{|A|}$, while the fraction on the right is the number of maximal chains in segment $[P,P^{\top}]$, isomorphic to $\mathcal P^{|P|}$. Here $\mathcal P^k$
denotes the lattice of partitions of a $k$-set. Thus for an atom
$[ij]$ and a partition $P\not\geqslant[ij]$, the number $\kappa_P^{[ij]}$ of maximal chains meeting both $P$ and $P\vee[ij]$ is $\kappa_P^{[ij]}=$
\begin{eqnarray*}
&=&\left[\prod_{1\leq k\leq n}\left(\frac{k!(k-1)!}{2^{k-1}}\right)^{c^P_k}\right]\frac{|P\vee[ij]|!(|P\vee[ij]|-1)!}{2^{|P\vee[ij]|-1}}\\
&=&\left[\prod_{1\leq k\leq n}(k!(k-1)!)^{c^P_k}\right]\frac{(|P|-1)!(|P|-2)!}{2^{|P|-2+\sum_{1\leq k\leq n}c^P_k(k-1)}}\\
&=&\left[\prod_{1\leq k\leq n}(k!(k-1)!)^{c^P_k}\right]\frac{(|P|-1)!(|P|-2)!}{2^{n-2}}\text ,
\end{eqnarray*}
yielding
\begin{equation*}
\frac{\kappa_P^{[ij]}}{\kappa}=2\left[\prod_{1\leq k\leq n}(k!(k-1)!)^{c^P_k}\right]\frac{(|P|-1)!(|P|-2)!}{n!(n-1)!}\text .
\end{equation*}
Therefore, the CU solution is
\begin{equation*}
\phi^{CU}_{[ij]}(f)=\sum_{P\not\geqslant[ij]}\frac{\kappa_P^{[ij]}}{\kappa}\left[\frac{f(P\vee[ij])-f(P)}{s(P\vee[ij])-s(P)}\right]\text ,
\end{equation*}
while the SU solution is, more simply,
\begin{equation*}
\phi^{SU}_{[ij]}(f)=\sum_{P\geqslant[ij]}\frac{\mu^f(P)}{s(P)}\text .
\end{equation*}
Now for $N=\{1,2,3\}$, consider the G game $f=\zeta_{[12]}$ given by basis element or zeta game $\zeta_{[12]}$ (see example 5). That is $\zeta_{[12]}(P)=1$ if $P\geqslant[12]$ and
0 otherwise, where $[12],[13]$ and $[23]$ are the three atoms. Then,%\footnote{It may be noted that all the five partitions in $\mathcal P^N,N=\{1,2,3\}$ are modular
%\cite{Aigner97,Stanley1971}, while as soon as $n>3$ non-modular partitions appear.}. Then,
\begin{eqnarray*}
\phi^{SU}_{[ij]}(\zeta_{[12]})&=&\left\{\begin{array}{c} 1\text{ if }[ij]=[12]\\0\text{ if }[ij]\neq[12]\end{array}\right.\text{ and }\\
\phi^{CU}_{[ij]}(\zeta_{[12]})&=&\left\{\begin{array}{c} 2/3\text{ if }[ij]=[12]\\1/6\text{ if }[ij]\neq[12]\end{array}\right.\text{, while}
\end{eqnarray*}
$\phi^{CU}_{[ij]}(r)=\phi^{SU}_{[ij]}(r)=2/3=\frac{r(P^{\top})}{\binom{3}{2}}=\phi_{[ij]}^{CU}(\zeta_{[ij]})$ for all $[ij]$.

The remainder of this Section is based on the approach to the geometric lattice of embedded subsets $(\mathcal E^N,\sqcap,\sqcup)$ defined in \cite{Rossi2017GeometricLattice}.
In particular, join $\sqcup$ obtains via a closure operator \cite{Aigner97} $cl:2^N\times\mathcal P^N\rightarrow2^N\times\mathcal P^N$ that cannot be detailed here for reasons of
space and mostly because it surely does not fit a contribution that is aimed to be useful in the area of game-theoretical models of cooperative communication networking. Clarified
this, if PFF games $h$ are looked at as functions defined on $\mathcal E^N$, with solutions $\phi(h)=\{\phi_a(h):a\in\mathcal E^N_{\mathcal A}\}\in\mathbb R^{\binom{n+1}{2}}$
being lattice functions themselves, then these latter have M\"obius inversion $\mu^{\phi(h)}(a)=\phi_a(h)$ living only on atoms $a\in\mathcal E^N_{\mathcal A}$, i.e.
$\phi(h)(A,P)=\sum_{a\sqsubseteq(A,P)}\phi_a(h)$ for all $(A,P)\in\mathcal E^N$.

Lattices $(\mathcal P^N,\wedge,\vee)$ and $(\mathcal E^N,\sqcap,\sqcup)$ are characterized by the following isomorphisms:\\
$[(\emptyset,P_{\bot}),(N,P^{\top})]=\mathcal E^n\cong\mathcal P^{n+1}$,\\
$[(\emptyset,P_{\bot}),(A,P)]\cong\mathcal E^{|A|}\times\left(\times_{B\in P^{A^c}}\mathcal P^{|B|}\right)$ and\\
$[(A,P),(N,P^{\top})]\cong\mathcal P^{|P|}$ if $A\neq\emptyset$,\\
while $[(A,P),(N,P^{\top})]\cong\mathcal E^{|P|}$ if $A=\emptyset$,\\
where $P^{A^c}=\{A^c\cap B:P\ni B,\emptyset\neq A^c\cap B\}$ is the partition of $A^c\neq\emptyset$ induced by $P=\{B_1,\ldots,B_{|P|}\}$ and $\mathcal E^k$ is the geometric
lattice of embedded subsets of a $k$-set. Accordingly, the number of maximal chains in $\mathcal E^N$ (i.e.
from $(\emptyset,P_{\bot})$ to $(N,P^{\top})$) is $\kappa=\binom{n+1}{2}\binom{n}{2}\binom{n-1}{2}\cdots\binom{2}{2}=\frac{(n+1)!n!}{2^n}$, and the number of those meeting
$(A,P)\in\mathcal E^N$ is
\begin{equation*}
\frac{(|A|+1)!|A|!|P|!(|P|-1)!}{2^{|A|}2^{|P|-1}}\prod_{k=1}^{n-|A|}%\prod_{1\leq k\leq n-|A|}
\left(\frac{k!(k-1)!}{2^{k-1}}\right)^{c^{P^{A^c}}_k}%\frac{|P|!(|P|-1)!}{2^{|P|-1}}
\end{equation*}
if $A\neq\emptyset$, and
\begin{equation*}
\frac{(|P|+1)!|P|!}{2^{|P|}}\prod_{k=1}^{n}%\prod_{1\leq k\leq n}
\left(\frac{k!(k-1)!}{2^{k-1}}\right)^{c^P_k}
\end{equation*}
if $A=\emptyset$. For an atom $a\in\mathcal E^N_{\mathcal A}$ and an embedded subset $(A,P)\not\sqsupseteq a$, the number $\kappa^a_{(A,P)}$ of maximal chains meeting both
$(A,P)$ and the embedded subset $cl((A,P)\sqcup a)$ obtained as the (above-mentioned) join of $(A,P)$ and $a$ thus is
\begin{equation*}
\frac{(|A|+1)!|A|!(|P|-1)!(|P|-2)!}{2^{|A|}2^{|P|-2}}\prod_{k=1}^{n-|A|}%\prod_{1\leq k\leq n-|A|}
\left(\frac{k!(k-1)!}{2^{k-1}}\right)^{c^{P^{A^c}}_k}%\frac{(|P|-1)!(|P|-2)!}{2^{|P|-2}}=
\end{equation*}
\begin{equation*}
=%(|A|+1)!|A|!
\frac{(|A|+1)!|A|!(|P|-1)!(|P|-2)!}{2^{n-1}}
\prod_{k=1}^{n-|A|}%\prod_{1\leq k\leq n-|A|}
\left((k!(k-1)!\right)^{c^{P^{A^c}}_k}
\end{equation*}
if $A\neq\emptyset$, and
\begin{equation*}
\frac{|P|!(|P|-1)!}{2^{|P|-1}}\prod_{k=1}^{n}%\prod_{1\leq k\leq n}
\left(\frac{k!(k-1)!}{2^{k-1}}\right)^{c^P_k}
\end{equation*}
\begin{equation*}
=\frac{|P|!(|P|-1)!}{2^{n-1}}\prod_{k=1}^{n}%\prod_{1\leq k\leq n}
\left(k!(k-1)!\right)^{c^P_k}\text{ if }A=\emptyset\text .
\end{equation*}
This means that $\frac{\kappa^a_{(A,P)}}{\kappa}=$
\begin{equation*}
=\frac{2(|P|-1)!(|P|-2)!(|A|+1)!|A|!}{(n+1)!n!}\prod_{k=1}^{n-|A|}%\prod_{1\leq k\leq n-|A|}
\left(k!(k-1)!\right)^{c^{P^{A^c}}_k}
\end{equation*}
if $A\neq\emptyset$, while% $\frac{\kappa^a_{(A,P)}}{\kappa}=$
\begin{equation*}
\frac{\kappa^a_{(A,P)}}{\kappa}=\frac{2|P|!(|P|-1)!}{(n+1)!n!}\prod_{k=1}^{n}%\prod_{1\leq k\leq n}
\left(k!(k-1)!\right)^{c^P_k}
\text{ if }A=\emptyset\text .
\end{equation*}
Therefore, the CU solution of PFF game $h$ is $\phi_a^{CU}(h)=$
\begin{equation*}
=\sum_{(A,P)\not\sqsupseteq a}\frac{\kappa^a_{(A,P)}}{\kappa}
\left[\frac{h(cl((A,P)\sqcup a))-h(A,P)}{s(cl((A,P)\sqcup a))-s(A,P)}\right]
\end{equation*}
while the SU solution is
\begin{equation*}
\phi^{SU}_a(h)=\sum_{(A,P)\sqsupseteq a}\frac{\mu^h(A,P)}{s(A,P)}\text .
\end{equation*}

For $N=\{1,2\}$, consider the PFF game $h=\zeta_{(\emptyset,[12])}$ given by the basis element or zeta game $\zeta_{(\emptyset,[12])}$, i.e.
$\zeta_{(\emptyset,[12])}(A,P)=1$ if $(A,P)\sqsupseteq(\emptyset,[12])$ and 0 otherwise (for all $(A,P)\in\mathcal E^N$), and
where $(1,1|2)$ and $(2,1|2)$ and $(\emptyset,[12])=(\emptyset,12)$ are the three atoms, with vertical bar $|$ separting blocks. That is,
%\footnote{Again, the five embedded subsets in $\mathcal E^{\{1,2\}}$ are all modular \cite{Rossi2017GeometricLattice}.},
$1|2=P_{\bot}$ while $12=[12]=P^{\top}$, where
$(\emptyset,1|2)\sqsubset(1,1|2),(2,1|2),(\emptyset,12)\sqsubset(\{1,2\},12)$, with bottom $(\emptyset,P_{\bot})=(\emptyset,1|2)$ and top $(\{1,2\},12)=(N,P^{\top})$. Then the
corresponding SU and CU solutions are:
\begin{eqnarray*}
\phi^{SU}_a(\zeta_{(\emptyset,[12])})&=&\left\{\begin{array}{c} 1\text{ if }a=(\emptyset,[12])\\0\text{ if }a=(1,1|2)\text{ or }a=(2,1|2)\end{array}\right.\text ,\\
\phi^{CU}_a(\zeta_{(\emptyset,[12])})&=&\left\{\begin{array}{c} 2/3\text{ if }a=(\emptyset,[12])\\
1/6\text{ if }a=(1,1|2)\text{ or }a=(2,1|2)\end{array}\right.\text ,
\end{eqnarray*}
and $\phi^{CU}_a(r)=\phi^{SU}_a(r)=2/3=\frac{r(N,P^{\top})}{\binom{3}{2}}=\phi_a^{CU}(\zeta_a)$ for all atoms $a\in\mathcal E^N_{\mathcal A}$, with $r(A,P)=r(P)+\min\{|A|,1\}$
as the rank function $r:\mathcal E^N\rightarrow\mathbb Z_+$.

\section{The core and additivity}
The core $\mathcal C(v)$ of C games $v$ is often introduced as the set of point-valued solutions $\phi(v)$ that no coalition $A\in 2^N$ can block, meaning that under
sharing rule $\phi$ every $A$ must receive an amount of TU $\phi(v)(A)=\sum_{i\in A}\phi_i(v)\geq v(A)$, with equality for the grand coalition $A=N$. In fact, if 
$\phi(v)(A)<v(A)$, then why should coalition members $i\in A$
cooperate with non-members $j\in A^c$? Thus, formally, looking at the $n$ values of M\"obius inversion $\mu^{\phi(v)}$, namely at the $n$ shares
$(\phi_1(v),\ldots,\phi_n(v))\in\mathbb R^n$, the core is a convex and possibly empty subset $\mathcal C(v)\subset\mathbb R^n$. It is the main set-valued solution concept, and
the necessary and sufficient conditions for non-emptyness $\mathcal C(v)\neq\emptyset$ are the well-known `Shapley-Bondareva conditions' (see, for instance,
\cite[p. 210]{GameTheoryCommunicationb00k2012}). On the other hand, supermodularity (also termed `convexity' \cite{Shapley71}) of $v$, that is
$v(A\cup B)+v(A\cap B)\geq v(A)+v(B)$ for all $A,B\in 2^N$, is a sufficient but not necessary condition for non-emptyness $\mathcal C(v)\neq\emptyset$, and total positivity
$\mu^v(A)\geq 0$ for all $A\in 2^N$ entails supermodularity (see Section II). In fact, if $v$ is supermodular, then the extreme points of (convex polyhedron) $\mathcal C(v)$ are
those solutions obtained by rewarding each player $i\in N$ with marginal contribution $v(A_k\cup i)-v(A_k)$ where $A_k,0\leq k\leq n$ is a maximal chain in $2^N$, i.e.
$A_{k+1}=A_k\cup i$. In such a case, the Shapley value $\phi^{Sh}(v)$ defined by expression (1) is the center of the core, in that $\phi^{Sh}$ is the `uniform' convex combination
of the $n!$ solutions associated with maximal chains as just specified (see \cite{Shapley71}). In multi-agent systems, $\mathcal C(v)$ is important precisely because coalitions
have no incentive to oppose any worth/cost-sharing rule that is known to be in the core \cite{Sandholm+2006,Rahwan+2007,IEEEtransVeicTech2011,PartitionFormSpecSensIEEE2012}.

When all of this is to be translated in terms G and PFF games, a first conceptual observation is that every partition involves all players, thus saying that `a partition of
players can block a sharing criterion' (or, more generally, an outcome) does not seem to allow for a straightforward interpretation. In fact, partition function-based TU
games model situations where all players cooperate, in some way. Specifically, every partition formalizes a distinct form of global cooperation, although the bottom $P_{\bot}$
seems to unambiguously represent the case where everyone stands alone\footnote{See \cite{GilboaLehrer90GG} on how to deal with a bottom partition whose worth is $\neq 0$.}.
Furthermore, as already explained in Section III in terms of monotonicity, the coarsest partition $P^{\top}$ appears to hardly correspond to the `best form of cooperation',
especially insofar as technology leads cooperative communication networking to realize through clusters of nodes. On the other hand, if the possibilities given by the technological
infrastructure are suitably translated by means of a M\"obius inversion living only on a proper interval (or segment) of the lattice (see expression (4)), then $P^{\top}$ does
formalize the required overall agreement on how to distribute the cost of network maintainance (or equivalently the worth of its existence), toward optimal global functioning.
Despite these premises, still from a geometric perspective the cores $\mathcal C(f)\subset\mathbb R^{\binom{n}{2}},\mathcal C(h)\subset\mathbb R^{\binom{n+1}{2}}$ of G and PFF
games $f,h$ are well-defined in terms of the novel solution concept proposed here. Maintaining the above notation $(L^N,\wedge,\vee)$ for a lattice
$L^N\in\{\mathcal P^N,\mathcal E^N\}$ in order to deal with both these games at once, Definition 1 in Section III leads to obtain the core
$\mathcal C(f)\subset\mathbb R^{|L^N_{\mathcal A}|}$ of games $f:L^N\rightarrow\mathbb R$ as the possibly empty convex polyhedron consisting of those solutions $\phi(f)$ such that
$\phi(f)(x)=\sum_{a\leqslant x}\phi_a(f)\geq f(x)$ for all elements $x\in L^N$, with equality for the top $x^{\top}$. Now, concerning both (i) the analog of the necessary and sufficient Shapley-Bondareva conditions
for non-emptyness, and (ii) supermodularity as a sufficient but not necessary such condition, the main difference with respect to C games is due, once again, to linear
dependence, which characterizes geometric (non-distributive) lattices \cite{Aigner97,Whitney35}. In fact, while noticing that in the CU and SU solutions, respectively, 
marginal contributions and the values of M\"obius inversion appear divided by the size of lattice elements, it must be considered that the size is a totally positive lattice function (as
its M\"obius inversion takes value 1 on atoms and 0 elsewhere). Therefore, in order for $\mathcal C(f)$ to be non-empty, it is not sufficient that $f$ is supermodular, i.e.
$f(x\vee y)+f(x\wedge y)\geq f(x)+f(y)$ for all $x,y\in L^N$. Conversely, $f$ has to quantify synergies minimally as great as those quantified by the size itself.
\begin{example}
Again, for $N=\{1,2,3\}$ consider the supermodular symmetric G game $f:\mathcal P^N\rightarrow\mathbb R$ defined by $f(P_{\bot})=0$, $f([ij])=1$ for $1\leq i<j\leq 3$ and
$f(P^{\top})=2$, where $f(P\vee Q)+f(P\wedge Q)\geq f(P)+f(Q)$ for all $P,Q\in\mathcal P^N$ is easily checked: if $P\geqslant Q$, then equality holds, while the
only remaining case is when both $P,Q\in\mathcal P^N_{\mathcal A}$ are atoms, but $f([12]\vee f([13]))+f([12]\wedge[13])=f([12])+f([13])=2$ (for instance), i.e. equality
holds as well. However, $f$ is \textit{not} totally positive: $\mu^f(P^{\top})=2-1-1-1=-1<0$, and in particular $f(P^{\top})<s(P^{\top})=3$. Thus $\mathcal C(f)=\emptyset$, as
$\phi_{[12]}(f),\phi_{[13]}(f),\phi_{[23]}(f)$ cannot satisfy both $\phi_{[ij]}(f)\geq 1$, $1\leq i<j\leq 3$ and
$\phi_{[12]}(f)+\phi_{[13]}(f)+\phi_{[23]}(f)=2$.
\end{example}

The necessary and sufficient conditions for non-emptyness of the core of G and PFF games deserve separate treatment.

\subsection{Additively separable partition functions}
Partition functions $f:\mathcal P^N\rightarrow\mathbb R$ are additively separable if a set function $v:2^N\rightarrow\mathbb R$ satisfies $f(P)=\sum_{A\in P}v(A)$ for all
$P\in\mathcal P^N$ \cite{GilboaLehrer90GG,GilboaLehrer91VI}, and are also sometimes termed, more simply, `additive'. They appear in a variety of settings, ranging from
combinatorial optimization problems \cite{KorteVygen2002} to community/module detection in complex networks \cite{FortunatoLongSurvey2010}. These partition functions admit in fact
a continuum of additively separating set functions, as exempified hereafter.
\begin{example}{\textsl{Additively separating the rank.}}
The rank of partitions $r(P)=n-|P|=\sum_{A\in P}(|A|-1)$ is immediately seen to be additively separated by the symmetric set function $v$ (see Subsection III.A) defined by
$v(A)=|A|-1$ for all $A\in 2^N$, whose M\"obius inversion takes values $\mu^v(\emptyset)=-1$, $\mu^v(A)=1$ if $|A|=1$ and $\mu^v(A)=0$ if $|A|>1$. However, $r(\cdot)$ may be
checked to be also additively separated by the (again symmetric) set function $v'$ with M\"obius inversion $\mu^{v'}(A)=0$ if $|A|\leq 1$ and $\mu^{v'}(A)=(-1)^{|A|}$ if $|A|>1$,
thus $v'(A)=0$ if $|A|\leq 1$ and $v'(A)=1$ if $|A|=2$, $v'(A)=2$ if $|A|=3$, $v'(A)=3$ if $|A|=4$, i.e. $\mu^{v'}(A)=|A|-1-\sum_{B\subset A}\mu^{v'}(B)$.
\end{example}
\begin{example}{\textsl{Additively separating the size.}}
The size of partitions $s(P)=\sum_{A\in P}\binom{|A|}{2}$ is immediately seen to be additively separated by the symmetric set function $v$ defined by
$v(A)=\binom{|A|}{2}$ for all $A\in 2^N$, whose M\"obius inversion takes values $\mu^v(A)=0$ if $|A|\neq 2$ and $\mu^v(A)=1$ if $|A|=2$. However, $s(\cdot)$ is
also additively separated by (symmetric) $v'$ with M\"obius inversion
$\mu^{v'}(A)=\left\{\begin{array}{c}(-1)^{|A|+1}\text{ if }|A|\neq 2\\0\text{ if }|A|=2\end{array}\right.$, hence 
$\mu^{v'}(\emptyset)=-1$, $\mu^{v'}(A)=1$ if $|A|=1$, $\mu^{v'}(A)=0$ if $|A|=2$, $\mu^{v'}(A)=1$ if $|A|=3$,
$\mu^{v'}(A)=-1$ if $|A|=4$, $\mu^{v'}(A)=1$ if $|A|=5$, $\mu^{v'}(A)=-1$ if $|A|=6$ and so on according to recursion $\mu^{v'}(A)=\binom{|A|}{2}-\sum_{B\subset A}\mu^{v'}(B)$.
\end{example}
In order to briefly generalize these examples, let $\mathbf{AS}(f)\subset\mathbb R^{2^n}$ denote the convex and possibly empty set\footnote{$v(\emptyset)=\mu^v(\emptyset)$
being chosen arbitrarily, any $\mathbf{AS}(f)\neq\emptyset$ is not bounded.} of set functions $v$ that additively separate any given partition
function $f$, i.e. $v,v'\in\mathbf{AS}(f),\alpha\in[0,1]\Rightarrow[\alpha v+(1-\alpha)v']\in\mathbf{AS}(f)$, as
$$\sum_{A\in P}[\alpha v(A)+(1-\alpha)v'(A)]=\alpha\sum_{A\in P}v(A)+(1-\alpha)\sum_{A\in P}v'(A)$$
$=\sum_{A\in P}v(A)=\sum_{A\in P}v'(A)$ for all $P\in\mathcal P^N$. Emptyness
$\mathbf{AS}(f)=\emptyset$ corresponds to a non-additively separable $f$. For $v\in\mathbf{AS}(f)$,
any $v'\in\mathbf{AS}(f)$ obtains recursively as follows:\smallskip\\
(i) $n\mu^{v'}(\emptyset)+\sum_{i\in N}\mu^{v'}(\{i\})=n\mu^v(\emptyset)+\sum_{i\in N}\mu^v(\{i\})$,\smallskip\\
entailing $\sum_{i\in N}v'(\{i\})=f(P_{\bot})=\sum_{i\in N}v(\{i\})$;\smallskip\\
(ii) $\mu^{v'}(A)=v(A)-\sum_{B\subset A}\mu^{v'}(B)$ for all $A,|A|>1$,\smallskip\\
entailing $v'(A)=v(A)$ for all $A,|A|>1$. A similar argument applies to the set $\mathbf{AS}(h)$ of set functions $v$ additively separating PFF games $h$, i.e. such that
$h(A,P)=v(A)+\sum_{B\in P}v(B)$ for all embedded subsets $(A,P)\in\mathcal E^N$ \cite{Rossi2017GeometricLattice}.

Solutions of additively separable G and PFF games can be approached as traditional solutions of C games (i.e. via the Shapley value in expression (1), see \cite{GilboaLehrer90GG,Myerson88}). But if $\mathbf{AS}(f)\neq\emptyset\neq\mathbf{AS}(h)$, then the CU and SU solutions proposed above provide novel criteria for distributing either the generated TU or else the costs of system maintainance.

\section{Conclusion}
In a time-varying communication network $G^t=(N,E^t)$, any pair $\{i,j\}\in N_2$ of nodes may be an edge, i.e. a vehicle for exchanging information, over some period $\Delta t>0$,
and if an edge has never appeared over the whole history (up to some `present' $t$), then it is simply to be regarded as one that has made no contribution (thus far). In this
view, each pair of nodes receives a (possibly null) share of the surplus generated by cooperation. Thus the idea is that exhaustive information about network topology and traffic
is constantly collected, in a distributed and local manner, enabling to use such an information for each period $t-1\rightarrow t$ in order to reward nodes at time $t$. A constant node set $N=\{1,\ldots ,n\}$ is also recognized to be employed here for expositional convenience, as not only at the begininning of each period a new
game starts being played, but new nodes may also enter at any time $t$.

In order to exemplify what form a fixed point of the SU solution of G games may take, consider the case where the surplus generated by cooperation is simply assumed to be
quantified by the volume of data traffic over the whole (clustered) network during each time period. In particular, if the protocol requires redundant transmissions, let all
transmissions contribute to the generated surplus independently from redundancy. In other terms, the (periodical) worth of global cooperation is the sum over all edges of the
traffic that occurs through them. Then, the share assigned to each edge by the SU solution is precisely the volume of data that is transferred through that edge, and thus such a
G game is a fixed point of the SU solution (mapping). In practice, this would mean that an arbitrary unit of data transfer (say 1 MB) is equivalent to an arbitrary unit of TU (say
a token giving temporary priviliges).

Concerning how to divide the share of each edge between its endnodes, in graph-restricted C games \cite{Myerson77Graph,Owen86} players must be
rewarded by taking into account only their position in the network and their marginal contributions to coalitions. In other terms, such players do not have different roles for
network functioning, and thus the share of edges seem best equally divided between endnodes (as suggested in \cite{Borm++92} for `communication situations'). However, in many
networking systems different nodes may well play different roles depending on their (time-varying) energy constraints and/or computational capabilities. In the simplest case, some
of them are cluster heads while the remaining ones are not. Similarly, in cognitive radio networks players may be either primary users or else secondary users. This suggests that
in several communication networking systems the shares distributed over edges can be next divided between nodes in more sophisticated manners than just via equal splitting. 

A final but seemingly very important remark concerns M\"obius inversion, which was defined to provide \textit{``the combinatorial analog of the fundamental theorem of calculus''}
\cite{Rota64}. Roughly speaking, M\"obius inversion may be regarded as the `derivative' of lattice (or more generally poset) functions, and constitutes a very useful tool for
game-theoretical modeling. For instance, apart from G and PFF games, several communication networking systems can be modeled by means of graph-restricted C games
\cite{GameTheoryCommunicationb00k2012,CoalitionalGameThoeryTutorial}, and in particular the Myerson value \cite{Myerson77Graph} appears to be an important solution for such
settings. In fact, as outlined in \cite[Sec. 7.5.2]{GameTheoryCommunicationb00k2012} and \cite[p. 23]{CoalitionalGameThoeryTutorial}, the Myerson value is well-known to be the
Shapley value (given by expression (1)) of a novel (i.e. graph-restricted) C game. However, it may be worth emphasizing that such a novel game $v/G$ is characterized as follows:
(i) it coincides with the original (unrestricted) C game $v$ on those coalitions spanning (or inducing \cite{DiestelGraphTheory2010}) a connected subgraph, and (ii) its M\"obius
inversion $\mu^{v/G}$ takes value 0 on the remaining coalitions \cite{Owen86}. Hence graph-restricted C game $v/G$ (also termed `coalitional graph game' in
\cite{GameTheoryCommunicationb00k2012,CoalitionalGameThoeryTutorial}) has M\"obius inversion $\mu^{v/G}$ living only on connected coalitions. Insofar as applications are concerned,
this means that the Myerson value may be computed by means of the second equality in expression (1), once the non-zero values of $\mu^{v/G}$ are recursively determined.
Hopefully, the present paper may thus contribute to fostering the use of M\"obius inversion (of all types of TU cooperative games) for modeling collaborative
communication systems.   

\subsection{Future work}
Definition 1 in Section III enables to reconsider the whole theory of C games in terms of G and PFF games. As for the core (see Section V), the well-known Shapley-Bondareva non-emptiness conditions may be paralleled with focus on what changes must be introduced in order to take into account the size of lattice elements. In this view, it seems that dividing the main expressions for the CU and SU solutions by the size (see above) is already in compliance with the comprehensive approach (to the core of C games) relying on `concavification' \cite{Concavification}. From a more general perspective, C games are commonly looked at as pseudo-Boolean functions, thereby obtaining the Shapley, Banzhaf and other values through the gradient of the polynomial multilinear extension of such functions \cite{Banzhaf88HolzmanLehrer}, \cite[pp. 139-151]{Roth88}. A challenging task thus is to reproduce the same entire pseudo-Boolean framework for G and PFF games.

\bibliographystyle{abbrv}
\bibliography{BIBLIO-PartitionFunctionCooperativeCommunication}

\vspace{12pt}
%\color{red}
%IEEE conference templates contain guidance text for composing and formatting conference papers. Please ensure that all template text is removed from your conference paper prior to submission to the conference. Failure to remove the template text from your paper may result in your paper not being published.

\end{document}